\documentclass[12pt,prd]{revtex4}

\usepackage{amsmath}
\usepackage{amssymb}
\usepackage[dvips]{graphicx,color}

\begin{document}

\title{
Fermion Actions extracted from 
\\ Lattice Super Yang-Mills Theories
}

\author{Tatsuhiro Misumi}
\email{misumi@phys-h.keio.ac.jp}
\affiliation{Department of Physics, and Research and Education Center for Natural Sciences, 
Keio University, Hiyoshi 4-1-1, Yokohama, Kanagawa 223-8521, Japan}

\begin{abstract}
We revisit 2D $\mathcal{N}=(2,2)$ super Yang-Mills lattice formulation (Sugino model)
to investigate its fermion action with two (Majorana) fermion flavors and exact chiral-$U(1)_{R}$ symmetry.
We show that the reconcilement of chiral symmetry and absence of further species-doubling
originates in the 4D clifford algebra structure of the action, where 2D two flavors are spuriously treated 
as a single 4D four-spinor with four 4D gamma matrices introduced into kinetic and Wilson terms.
This fermion construction based on the higher-dimensional clifford algebra 
is extended to four dimensions in two manners:
(1) pseudo-8D sixteen-spinor treatment of 4D four flavors with eight 8D gamma matrices,
(2) pseudo-6D eight-spinor treatment of 4D two flavors with five out of six 6D gamma matrices. 
We obtain 4D four-species and two-species lattice fermions with unbroken subgroup of chiral symmetry
and other essential properties.
We discuss their relations to staggered and Wilson twisted-mass fermions.
We also discuss their potential feedback to 4D super Yang-Mills lattice formulations.
\end{abstract}

\maketitle

\newpage


\section{Introduction}
\label{sec:Intro}

For the purpose of realizing efficient lattice computations and extending 
its applicability to a broader range of fields, development of better lattice fermion discretizations 
is one of the most important tasks.
All the known lattice fermion constructions to circumvent the no-go theorem 
\cite{KarS, NN} have their individual shortcomings:  
Wilson fermions with explicit chiral symmetry breaking and $O(a)$ discretization errors,
domain-wall \cite{Kap, Fu1, FuSh} and overlap \cite{GW, Neu} fermions with expensive simulation algorithms,
and staggered fermions  \cite{KS, Suss, KaS, Sha, GS} with four tastes and $O(a^2)$ taste-breaking.
To date, several skillful methods for reducing errors and improving chiral properties 
have been developed both for Wilson and staggered fermions, including the smeared-link clover \cite{DHK1}, 
Wilson twisted-mass \cite{F1, FR1} and HISQ \cite{Fo1}, etc.
On the other hand, we may still have unknown fermion setups which drastically improve lattice computations.

Recent progress in lattice fermion formulations has been made in the following topics:
Minimally doubled fermions \cite{Kar, Wilc, Creutz1, Bori, CM, M1} 
with the minimal number of species, or two species, possesses exact chiral symmetry
and ultra locality, while its practical application requires fine-tuning of at least two parameters because of
the explicit breaking of hypercubic and C, P, T invariance \cite{Bed1, Bed2, Cap1, Cap2, Cap3, Kama1}.
Taste-splitting staggered fermion \cite{Adams1} is 
a staggered-fermion version of Wilson or overlap formulations \cite{Adams2, Hoel, PdF, CKM2},
which has possibilities for reducing numerical costs with lattice QCD with overlap setups \cite{Durr-StW}.
Flavored-mass and Brillouin fermions \cite{CKM1, Durr1, hi} are extensions of Wilson fermion
with different species-splitting manners, 
which can be a better setup for heavy quark systems as well as a better kernel for overlap fermions \cite{Durr2}.
For the recent progress on fermion discretizations, see \cite{M2} for example.

In the present work, we search for a new idea of lattice fermion setups in
the supersymmetric lattice formulation \cite{KKU1, CKKU1,CKKU2, KU1, Su1, Su2, Su3, Su4, Su5}.
The reason why we here pay attention to SUSY lattice models is as follows:
Since an influence of species-doubling is also propagated to a bosonic part in the supersymmetric theory,
the number of species must be under control strictly in the associated lattice model.
Furthermore, chiral R symmetry is an essential element 
in the extended supersymmetric theories and it must be unbroken in the associated lattice models.
To sum up, if there is a well-defined (extended) supersymmetric lattice setup,
it is strongly expected that the lattice model automatically settles species-doubling 
with keeping subgroup of exact chiral symmetry in some way.
Some of the successful supersymmetric lattice theories adopt domain-wall and overlap formulations \cite{SUSY2}
as fermion parts, while others may realize reconcilement of chiral symmetry and absence of doublers 
in some unknown way. 
This is the point where a new idea for bypassing the doubling problem could be hidden.

In this paper, we investigate the two-dimensional $\mathcal{N}=(2,2)$ supersymmetric 
Yang-Mills (SYM) lattice theory invented in \cite{Su1,Su2, Su3}, 
with emphasis on the fermion action containing two Majorana d.o.f.s
and possessing exact chiral-$U(1)$ R symmetry.
(See \cite{SUSY3} for details of the model and the related numerical simulations.)
We find out that the reconcilement of exact chiral symmetry and the absence of further doublers
originates in 4D clifford algebra structure embedded into the 2D action, 
where 2D two flavors are spuriously treated as a single 4D four-spinor 
and four 4D gamma matrices are introduced into kinetic and Wilson terms.
This structure works to avoid further species-doubling and protect chiral-$U(1)$ R symmetry.
In other words, the setup identifies two species as two fermion flavors associated with the $\mathcal{N}=(2,2)$ 
extended supersymmetry, which is related to 4D $\mathcal{N}=1$ SUSY via dimensional reduction.

We extend this fermion construction based on the higher-dimensional clifford algebra 
to lattice Dirac fermions in four dimensions in two different manners:
In the first formulation, a 4D four-flavor multiplet is treated as a single pseudo 8D sixteen-spinor field, 
with eight elements of 8D gamma matrices introduced to the action. 
Due to the 8D clifford algebra, no further doublers emerge while subgroup of exact chiral symmetry remains unbroken.
It has several common symmetries with staggered fermions,
but has smaller translation invariance and larger flavor-breaking at $O(a)$. 
We discuss its application to 4D $\mathcal{N}=4$ SYM lattice theory.
In the second setup, two flavors are assembled into a single pseudo-6D eight-spinor field, 
and five of six 6D gamma matrices are assigned to kinetic and Wilson terms. 
We obtain the lattice fermion action with two species, exact chiral symmetries, hypercubic symmetry
and flavored-P,T invariances, which is shown to be equivalent to the Wilson twisted-mass fermion.
We discuss possibility of its application to 4D $\mathcal{N}=2$ SYM lattice theory.

In Sec.~\ref{sec:SUSY} we investigate the structure of fermion action in lattice formulation of
two-dimensional $\mathcal{N}=(2,2)$ super Yang-Mills theory.
In Sec.~\ref{sec:2D} we construct the 2D two-flavor lattice Dirac fermion
based on the four-dimensional clifford algebra.
In Sec.~\ref{sec:4D4} we propose the 4D four-flavor lattice fermion 
based on the eight-dimensional clifford algebra and discuss its symmetry.
In Sec.~\ref{sec:4D} we construct the 4D two-flavor lattice fermion,
and discuss its equivalence to the Wilson twisted-mass fermion.
Section \ref{sec:SUM} is devoted to a summary.


\section{Fermion Action in $\mathcal{N}=(2,2)$ SUSY lattice}
\label{sec:SUSY}

In this section we investigate the fermion action of the two-dimensional $\mathcal{N}=(2,2)$ 
super Yang-Mills lattice formulation.

The lattice model for 2D $\mathcal{N}=(2,2)$ super Yang-Mills theory invented by Sugino \cite{Su1,Su2,Su3,Su4,Su5}
is based on the topological-twisted form of the action \cite{W1,W2, Cat1}:
The original continuum $\mathcal{N}=(2,2)$ theory has four real supercharges $Q_{\pm}$, $\bar{Q}_{\pm}$ and four real
one-spinors (two Majorana spinors) $\lambda_{\pm}$, $\bar{\lambda}_{\pm}$ with two $U(1)$ R-symmetries we denote as $U(1)_{R}\times U(1)_{L}$.
In the topological twist, the subscripts associated with one of R-rotation $U(1)_{L}\sim SO(2)_{L}$ and Euclidian rotation $SO(2)_{E}$ are mixed, or equivalently, we consider diagonal subgroup $SO(2)_{L'}$ of $SO(2)_{L} \otimes SO(2)_{E}$ to redefine quantum number.
On the flat spacetime, this procedure is just change of field variables, 
and the twisted theory is equivalent to the original one.
In this basis we have a scalar supercharge (BRST charge) $Q$ as a linear combination of the supercharges,
which is nilpotent $Q^{2}=0$ up to gauge transformation.
Sugino's lattice model is constructed to keep invariance under this scalar supercharge $Q$,
which is not related with the infinitesimal translation \cite{SUSY1} and can be seen as fermionic internal symmetry.
Indeed, the action in the Sugino model \cite{Su1} is written in a $Q$-exact form.
We note that, in this basis, the four real one-spinors $\lambda_{\pm}$, $\bar{\lambda}_{\pm}$ 
are also transformed into topological-twisted forms including 
a 0-form $\eta$, a one-form $\psi_{\mu}$  and a two-form $\chi_{\mu\nu}=\chi$ ($\mu,\nu=1,2$). 
 
Now, we look into the fermion action part of the $\mathcal{N}=(2,2)$ lattice formulation \cite{Su1,Su2,Su3}.
Instead of the standard two-spinor fermion action in two dimensions,
the fermion variable is introduced as a single four-spinor $\Psi=(\psi_{1},\psi_{2},\chi,{1\over{2}}\eta)^{T}$,
where the original d.o.f. of two 2D Majorana flavors is assembled into a single pseudo-4D fermion field.
Such four-spinor treatment is quite natural since 
2D $\mathcal{N}=(2,2)$ super Yang-Mills theory is obtained from
4D $\mathcal{N}=1$ super Yang-Mills theory through the dimensional reduction.
The action is given by
\begin{align}
S_{F}^{\mathcal{N}=(2,2)}=-{a^{3}\over{4g_{0}^{2}}}\sum_{n,\mu}{\rm tr}\,
\Psi_{n}^{T}
[\Gamma_{\mu}(U_{n,\mu}\Psi_{n+\mu}-U_{n,-\mu}\Psi_{n-\mu})
+P_{\mu}(2-U_{n,\mu}\Psi_{n+\mu}-U_{n,-\mu}\Psi_{n-\mu})]\,,
\label{SuS}
\end{align}
with a bare gauge coupling $g_{0}$, lattice spacing $a$, two-dimensional lattice sites $n=(n_{1}, n_{2})$
and a link variable $U_{n,\mu}$.
The trace ${\rm tr}$ is taken since fermions are in the adjoint representation.
$\Gamma_{\mu}$ and $P_{\mu}$ ($\mu=1,2$) belong to the four-dimensional gamma matrices.
We note that the first and second terms are kinetic and Wilson-like terms respectively.
As explicit forms of these matrices, the reference \cite{Su1} chooses the following,
\begin{equation}
\Gamma_{1}
=-i
\left(
\begin{array}{cc}
 & \sigma_{1}  \\
\sigma_{1} &   \\
\end{array}
\right)\,,\,\,\,\,\,\,\,
\Gamma_{2}
=i
\left(
\begin{array}{cc}
 & \sigma_{3}  \\
\sigma_{3} &   \\
\end{array}
\right)\,,\,\,\,\,\,\,\,
P_{1}
=
\left(
\begin{array}{cc}
& \sigma_{2}  \\
\sigma_{2} &  \\
\end{array}
\right)\,,\,\,\,\,\,\,
P_{2}
=i
\left(
\begin{array}{cc}
& -{\bf 1}  \\
{\bf 1} &  \\
\end{array}
\right)\,.
\label{SuG}
\end{equation}
with Pauli matrices $\sigma_{j}$ ($j=1,2,3$).
These four gamma matrices satisfy the four-dimensional clifford algebra as
\begin{equation}
\{ \Gamma_{\mu},\, \Gamma_{\nu}  \}=-2\delta_{\mu\nu}\,,\,\,\,\,\,\,\,
\{ P_{\mu},\, P_{\nu} \} = 2\delta_{\mu\nu}\,,\,\,\,\,\,\,\,
\{\Gamma_{\mu},\, P_{\nu}\}=0\,,
\label{4cli}
\end{equation}
where the minus sign in the first relation just arises from the anti-hermitian definition of
$\Gamma_{\mu}$, which can be translated into hermitian definition 
as $\Gamma_{\mu}\leftrightarrow i\Gamma_{\mu}$.

An outstanding point of this fermion action is
that, due to the 4D clifford algebra (\ref{4cli}), it prevents ``further" species-doubling from developing.
The associated free Dirac operator in the momentum space is given by
\begin{equation}
aD(p)= \sum_{\mu=1}^{2}\left[
-i\,\Gamma_{\mu}\,\sin ap_{\mu} + P_{\mu}\,(1-\cos ap_{\mu})
\right]\,.
\end{equation}
The number of species is counted by looking into zeros of the denominator of the Dirac propagator, 
which we denote as $|a^{2}D^{2}(p)|$,
\begin{equation}
|a^2D^{2}(p)|=\sum_{\mu=1}^{2}\left[
\sin^{2}ap_{\mu}+(1-\cos ap_{\mu})^{2}
\right]\,.
\end{equation}
Because of the 4D clifford algebra (\ref{4cli}),
this operator has no contact term between $\sin$ (kinetic) and $\cos$ (Wilson) terms.
Hence, it is clear that the number of zeros is only one at $ap=(0,0)$,
and the action has no further species-doubling.

This action possesses the exact chiral-$U(1)_{R}$ symmetry.
The Dirac operator $D(p)$ anti-commutes with the four product of
$\Gamma_{1},\Gamma_{2}, P_{1}$ and $P_{2}$
since the action contains $\Gamma_{\mu}$ in the kinetic term and $P_{\mu}$ in the Wilson term.
We denote the generator as
\begin{equation}
\Gamma_{5}=\Gamma_{1}\Gamma_{2}P_{1}P_{2}\,, 
\label{Hchi}
\end{equation}
which satisfies the relation
\begin{equation}
\{ D(p),\,\Gamma_{5}  \} = 0\,,
\end{equation}
It means that the fermion action (\ref{SuS}) is invariant under the following transformation,
\begin{align}
&\Psi_{n}\,\,\,\,\to\,\,\,\,\exp[i\,\theta\, \Gamma_{5}]\Psi_{n}\,,\,\,\,\,\,\,\,\,
\\
&\Psi^{T}_{n}\,\,\,\,\to\,\,\,\,\Psi^{T}_{n}\exp[i\,\theta\,\Gamma_{5}]\,.
\end{align}
This transformation is the chiral $U(1)_{R}$ rotation, 
which gives rotation in the multiple supercharge space, or the ``flavor" space.
We note that this chiral symmetry is realized as the pseudo higher-dimensional chiral 
symmetry in this formulation (\ref{Hchi}). 
For the explicit forms of gamma matrices (\ref{SuG}), 
the generator
\begin{equation}
\Gamma_{5}=\Gamma_{1}\Gamma_{2}P_{1}P_{2}=
\left(
\begin{array}{cc}
 {\bf 1} &   \\
 & -{\bf 1}   \\
\end{array}
\right)\,,
\end{equation}
rotates phases of $\psi_{1}, \psi_{2}$ and $\chi, \eta$ in opposite directions.
This exact invariance under the chiral-$U(1)_{R}$ prohibits mass renormalization of the fermion field 
in the model of 2D $\mathcal{N}=(2,2)$ super Yang-Mills theory.

One may have a question how the doubler-free lattice fermion can have exact chiral symmetry.
As we noted, the action originally contains two Majorana flavors in the four-spinor fermion field.
In the original paper \cite{Su1}, it is argued that the reason of the absence of doubling in the action 
is the explicit breaking of $U(1)$ fermion number (baryon) symmetry.
However, we note that, although the loss of $U(1)$ baryon invariance 
breaks an assumption of the no-go theorem \cite{NN},
it does not directly mean absence of unnecessary fermionic modes 
as seen in the surplus modes in 4D Sugino models \cite{Su3}. 
Furthermore, the breaking of $U(1)$ baryon invariance is just specific to the present form of the action,
and we still have the same property in the Dirac fermion version which possesses exact $U(1)$ baryon symmetry, 
as will be shown in the next section.
We argue that
{\it the main reason why the chiral-symmetric fermion action of $\mathcal{N}=(2,2)$ lattice model has no unnecessary fermionic modes is that it has 4D clifford-algebra structure, where two degrees of freedom of 2D fermions are treated as the pseudo-4D four-spinor fermion $\Psi$.} 
In the next section, we show details of the present formulation 
by studying the case of Dirac fermion action.


\section{2D two-flavor Dirac fermion}
\label{sec:2D}

In this section we consider the Dirac fermion version of the Sugino fermion action (\ref{SuS})
in order to clarify the argument in the previous section and also extend the idea to
more general cases.

We begin with the following 2D two-flavor Dirac fermion action on the lattice
\begin{equation}
S={1\over{2}}\sum_{n}\sum_{\mu=1}^{2}\bar{\Psi}_{n}\left[  
\Gamma_{\mu}(U_{n,\mu}\Psi_{n+\mu}-U_{n,-\mu}\Psi_{n-\mu})
+iP_{\mu}(2-U_{n,\mu}\Psi_{n+\mu}-U_{n,-\mu}\Psi_{n-\mu})
+2m_{0}\Psi_{n}
\right]\,,
\label{2DS}
\end{equation}
where we introduce two-flavor fermion fields $\Psi=(\psi_{A}, \psi_{B})^{T}$, 
$\bar{\Psi}=(\bar{\psi_{A}}, \bar{\psi_{B}})$ assembled into
a single pseudo-4D four-spinor field. (each of $\psi_{A}$, $\psi_{B}$, 
$\bar{\psi}_{A}$, $\bar{\psi}_{B}$ are usual 2D Dirac fermion fields.)
The second term in (\ref{2DS}) is the Wilson-like term accompanied by $P_{\mu}$.
$\Gamma_{\mu}$ and $P_{\mu}$ ($\mu=1,2$) belong to 
the four-dimensional gamma matrices $\gamma^{(4)}_{M}$ ($M=1,2,3,4$).
These matrices therefore satisfy the four-dimensional clifford algebra
\begin{equation}
\{ \Gamma_{\mu}, \Gamma_{\nu}  \}=2\delta_{\mu\nu}\,, \,\,\,\,\,\,\,
\{ P_{\mu}, P_{\nu} \} = 2\delta_{\mu\nu}\,,\,\,\,\,\,\,\,
\{\Gamma_{\mu},P_{\nu}\}=0\,.
\label{4cli2}
\end{equation}
In this case, we choose both $\Gamma_{\mu}$ and $P_{\mu}$ as hermitian.
The associated free Dirac operator in the momentum space in a massless theory is given by
\begin{equation}
D(p)=i\Gamma_{\mu}\sin p_{\mu}+iP_{\mu}(1-\cos p_{\mu})\,.
\label{2DfD}
\end{equation}
We here make all the fields and parameters dimensionless
as $a^{3/2}\psi \to \psi$, $ma\to m_{0}$, $ap_{\mu}\to p_{\mu}$ and $aD(p)\to D(p)$.

As explicit $\Gamma_{\mu}$ and $P_{\mu}$,
we, for example, take the following forms,
\begin{align}
\Gamma_{1}
&={\bf 1}\otimes \sigma_{1}
=
\left(
\begin{array}{cc}
\sigma_{1} &  \\
& \sigma_{1}   \\
\end{array}
\right)\,,\,\,\,\,\,\,\,\,\,
\Gamma_{2}=
{\bf 1}\otimes \sigma_{2}
=
\left(
\begin{array}{cc}
\sigma_{2}&   \\
 & \sigma_{2}   \\
\end{array}
\right)\,,
\nonumber\\
P_{1}
&=
-\sigma_{2}\otimes\sigma_{3}
=
\left(
\begin{array}{cc}
& i\sigma_{3}  \\
-i\sigma_{3} &  \\
\end{array}
\right)\,,\,\,\,\,\,\,\,\,
P_{2}
=
\sigma_{1}\otimes\sigma_{3}
=
\left(
\begin{array}{cc}
& \sigma_{3}  \\
\sigma_{3} &  \\
\end{array}
\right)\,.
\end{align}
Since these satisfies the four-dimensional clifford algebra (\ref{4cli2}),
we have only one zero at $p=(0,0,0,0)$ in the denominator of the Dirac propagator
\begin{equation}
|D(p)|^2=\sum_{\mu}^{2}\left[\sin^{2} p_{\mu}+(1-\cos p_{\mu})^{2}\right]\,.
\end{equation}

The Dirac propagator anti-commute with the product of $\Gamma_{\mu}$ and $P_{\mu}$ as
\begin{equation}
\Gamma_{5}=\Gamma_{1}\Gamma_{2}P_{1}P_{2}=(\sigma_{3}\otimes \sigma_{3})
=
\left(
\begin{array}{cc}
\gamma_{5}&   \\
 & -\gamma_{5}  \\
\end{array}
\right)\,,\,\,\,\,\,\,\,\,\,\,\,\,\,\,\,\,\,
\{ D(p),\,\Gamma_{5}  \} = 0\,,
\end{equation}
where we define the two-dimensional gamma-5 as $\sigma_{3}\equiv\gamma_{5}$.
This action is thus invariant under the following chiral rotation,
\begin{align}
&\Psi_{n}\,\,\,\,\to\,\,\,\,\exp[i\,\theta\, \Gamma_{5}]\Psi_{n}\,,\,\,\,\,\,\,\,\,
\\
&\bar{\Psi}_{n}\,\,\,\,\to\,\,\,\,\bar{\Psi}_{n}\exp[i\,\theta\,\Gamma_{5}]\,.
\end{align}
This is the $\sigma_{3}$-flavored subgroup of the $SU(2)$ axial symmetry.

A notable point in this action is that it breaks $SU(2)$ flavor symmetry at $O(a)$.
It is due to the Wilson-like term (\ref{2DS}), which causes flavor-Lorentz mixing.
This is natural since the present action is originally based on the topologically-twisted action
of $\mathcal{N}=(2,2)$ SYM, which mixes Lorentz rotation $SO(2)_{E}$ and flavor-R rotation 
$U(1)_{L}\sim SO(2)_{L}$.

One may be reminded of staggered fermions by this flavor-Lorentz mixing.
The Dirac operator (\ref{2DfD}) is rewritten as 
\begin{align}
D(p)&=i({\bf 1}\otimes \sigma_{\mu})\sin p_{\mu}+(\sigma_{\mu}\sigma_{3}\otimes \sigma_{3})(1-\cos p_{\mu})
\nonumber\\
&=i({\bf 1}\otimes \gamma_{\mu})\sin p_{\mu}+(\gamma_{\mu}\gamma_{5}\otimes \gamma_{5})(1-\cos p_{\mu})\,,
\label{PTB}
\end{align}
where we introduce a flavor-spin representation as ${\bf 1}\otimes \sigma_{\mu}$
and define $\sigma_{\mu}\equiv\gamma_{\mu}$ and $\sigma_{3}\equiv\gamma_{5}$.
This action or Dirac operator are incorrectly referred to as staggered fermions in most of textbooks \cite{MM1, Ro1}
However, it is clearly different from staggered fermions.
First of all, the flavor symmetry (taste) breaking starts from $O(a)$ in the present free action, 
while the staggered fermion just has $O(a^2)$ taste mixing due to the exchange of highly-virtual gluons.
The reason why this action is derived from the staggered fermion is well elucidated in \cite{La1}:
The form of the action (\ref{PTB}) is obtained from the co-ordinate space spin-flavor representation of staggered fermions.
The hypercubic blocking to construct the co-ordinate-space spin-flavor basis breaks translation invariance 
into double lattice-spacing translation invariance.
The final representation is, then, no longer the staggered fermion.
We note that such taste-breaking is not seen in the momentum-space spin-flavor representation in a free theory,
which does not break translation invariance \cite{GS}.
We may call the present fermion action Halved-translation-invariance (per dimension) 
fermion to distinguish from staggered fermions.

We note that our result is not inconsistent with the no-go theorem \cite{KarS, NN}.
The fermion field in the present setup originally contains two flavors, or two species, 
and our result is consistent with the no-go theorem since ``two" is the minimum number required by the theorem.
On the other hand, it breaks the conditions for the Kasten-Smit's lemma \cite{KarS}, which states that
we need to break hypercubic symmetry or reflection positivity in order to
reduce the number of doublers from $2^{D}$ to less. ($D$ stands for dimensions.)
In this case, hypercubic symmetry is broken into the discrete diagonal part $SW_{2}$ of
Lorentz $SO(2)$ and flavor $SU(2)$ rotations as with the staggered fermion.
We list up some relevant symmetries of the two-flavor fermion action (\ref{2DS}):
\vspace{2mm}

1. Flavor-non-singlet chiral symmetry with $\Gamma_{5}=\sigma_{3}\otimes \gamma_{5}$, 
\vspace{2mm}

2. $\gamma_{5}$-hermiticity with $\Gamma_{5}=\sigma_{3}\otimes \gamma_{5}$,
\vspace{2mm}

3. Discrete flavor-rotation symmetry $SW_{2}$,
\vspace{2mm}

4. Flavored-parity and -time-reversal invariances associated with $\sigma_{3}$,
\vspace{2mm}

5. $U(1)$ baryon symmetry.
\vspace{2mm}

Contrary to the fermion action in the Sugino model, 
the action is invariant under $U(1)$ fermion number invariance.
\begin{align}
&\Psi_{n}\,\,\to\,\,\exp[i\theta ({\bf 1}\otimes {\bf 1})]\Psi_{n},\,\,\,\,\,\,\,\,
\\
&\bar{\Psi}_{n}\,\,\to\,\,\bar{\Psi}_{n}\exp[-i\theta({\bf 1}\otimes {\bf 1})].
\end{align}
It indicates that the absence of doubler zeros does not arise from the breaking of
fermion number invariance. Regarding the flavored-P and flavored-T invariance,
we discuss details in the next section along with extension to four dimensions.

In the end of this section, we summarize the recipe for the 2D two-flavor lattice fermion setup
by use of the 4D clifford algebra.
\vspace{2mm}

(1) Identify two flavors of 2D two-spinors as a single pseudo-4D four-spinor field spuriously.
\vspace{2mm}

(2) Introduce four 4D gamma matrices and assign half to
a kinetic term as $\Gamma_{\mu}$ and the other half to a Wilson term as $P_{\mu}$.
\vspace{2mm}

(3) Due to the 4D clifford algebra, we can avoid further species-doubling
with keeping subgroup of exact chiral symmetry.


\section{Extension to 4D four-flavor fermion}
\label{sec:4D4}

In this section, we discuss the four-dimensional Dirac versions of the fermion construction 
by use of higher-dimensional clifford algebra.
A naive extension is by introducing the six-dimensional clifford algebra
into four-dimensional fermion actions to obtain the two-flavor fermion with exact chiral symmetry.  
Though, the six-dimensional clifford algebra consists of six elements of gamma matrices
$\gamma^{(6)}_{M}$ ($M=1,2,\cdot\cdot\cdot ,6$), which are insufficient for 
necessary eight anti-commuting matrices for $\Gamma_{\mu}$ and $P_{\mu}$ ($\mu=1,2,3,4$) 
in the present extension of the method in Sec.~\ref{sec:2D}.

In other words, if we have four $\Gamma_{\mu}$ and four $P_\mu$ with 
two flavors of 4D four-spinors identified as a pseudo-6D eight-spinor field,
they cannot satisfy the requisite anti-commuting relation,
\begin{equation}
\{ \Gamma_{\mu}, \Gamma_{\nu}  \}=2\delta_{\mu\nu}\,, \,\,\,\,\,\,\,
\{ P_{\mu}, P_{\nu} \} = 2\delta_{\mu\nu}\,,\,\,\,\,\,\,\,
\{\Gamma_{\mu},P_{\mu}\}\not=0\,,
\label{4cli-NA}
\end{equation}
with $\mu=1,2,3,4$.
It means the Dirac propagator can have poles other than $p=(0,0,0,0)$.
This is one reason why the $\mathcal{N}=2$ four-dimensional super Yang-Mills lattice formulation 
is not yet well-constructed as shown in \cite{Su3}.

However, we have several ways of the extension.
We discuss them in this and next sections.
One simple way of extending the idea to four dimensions is
to give up the two-flavor property, and instead, consider
four Dirac flavors and eight-dimensional clifford algebra. 
Then, the recipe is altered as follows,
\vspace{2mm}

(1) Treat four flavors of 4D four-spinors as a single pseudo-8D sixteen-spinor field.
\vspace{2mm}

(2) Introduce eight 8D gamma matrices and assign half to
a kinetic term as $\Gamma_{\mu}$ and the other half to a Wilson term as $P_{\mu}$.
\vspace{2mm}

(3) Due to the 8D clifford algebra, we can avoid further species-doubling
with keeping subgroup exact chiral symmetry.
\vspace{2mm}

The four-dimensional action in this way of extension is given by
\begin{equation}
S={1\over{2}}\sum_{n}\sum_{\mu=1}^{4}\bar{\Psi}_{n}\left[  
\Gamma_{\mu}(U_{n,\mu}\Psi_{n+\mu}-U_{n,-\mu}\Psi_{n-\mu})
+iP_{\mu}(2-U_{n,\mu}\Psi_{n+\mu}-U_{n,-\mu}\Psi_{n-\mu})
+2m_{0}\Psi_{n}
\right]\,,
\label{4D4S}
\end{equation}
where $\Psi=(\psi_{A}, \psi_{B}, \psi_{C}, \psi_{D})^{T}$ 
stands for a 4D four-flavor multiplet treated as a single pseudo-8D sixteen-spinor field.
$\Gamma_{\mu}$ and $P_{\mu}$ belong to the eight 8D gamma matrices 
$\gamma^{(8)}_{M}$ ($M=1,\cdot\cdot\cdot, 8$).
They satisfy the eight-dimensional clifford algebra as
\begin{equation}
\{ \Gamma_{\mu}, \Gamma_{\nu}  \}=2\delta_{\mu\nu}\,, \,\,\,\,\,\,\,
\{ P_{\mu}, P_{\nu} \} = 2\delta_{\mu\nu}\,,\,\,\,\,\,\,\,
\{\Gamma_{\mu},P_{\mu}\}=0\,.
\label{8cli}
\end{equation}
with $\mu=1,2,3,4$.
Due to this relation, the fermion action avoids further species-doubling.
Since the associated Dirac operator anti-commutes with the product of $\Gamma_{\mu}$
and $P_{\mu}$,
\begin{equation}
\{D,\Gamma_{5}\}=0\,,\,\,\,\,\,\,\,\,\,\,\,\,\,
\Gamma_{5}=\Gamma_{1}\Gamma_{2}\Gamma_{3}\Gamma_{4}P_{1}P_{2}P_{3}P_{4}\,,
\end{equation}
the fermion action is invariant under the following transformation
\begin{align}
&\Psi_{n}\,\,\,\,\to\,\,\,\,\exp[i\,\theta\, \Gamma_{5}]\Psi_{n}\,,\,\,\,\,\,\,\,\,
\\
&\bar{\Psi}_{n}\,\,\,\,\to\,\,\,\,\bar{\Psi}_{n}\exp[i\,\theta\,\Gamma_{5}]\,.
\end{align}

As explicit gamma matrices, for example, we consider the following choices,
\begin{align}
\Gamma_{\mu}&
={\bf 1}\otimes \gamma_{\mu}
=\left(
\begin{array}{cc}
\gamma_{\mu} &  \\
 & \gamma_{\mu}  \\
\end{array}
\right)\,\,\,\,\,\,\,\,\,\,(\mu=1,2,3,4)\,,
\nonumber\\
P_{\mu}&=-i\gamma_{\mu}\gamma_{5}\otimes\gamma_{5}\,,
\label{G4D4F}
\end{align} 
where we again introduce the flavor-spin representation and
$\gamma_{\mu}$, $\gamma_{5}$ stand for four-dimensional hermite gamma matrices.
Other seemingly-different choices of $P_{\mu}$, for example, $P_{\mu}=\gamma_{\mu}\otimes \gamma_{5}$,
can be derived by change of variables as $\bar{\Psi}\to\bar{\Psi}\exp[-i\pi/4 \, (\gamma_{5}\otimes {\bf 1})]$, 
$\Psi\to \exp[i\pi/4\, (\gamma_{5}\otimes {\bf 1})]\Psi$.

The action (\ref{4D4S}) with this basis gives the following free massless Dirac operator in a momentum space,
\begin{equation}
D(p)=i({\bf 1}\otimes \gamma_{\mu})\sin p_{\mu}+(\gamma_{\mu}\gamma_{5}\otimes \gamma_{5})(1-\cos p_{\mu})\,.
\label{PTB4D}
\end{equation}
This is a four-dimensional version of the Halved-translation-invariance fermion (\ref{PTB}),
which has been misunderstood as ``staggered fermions".
For the present choice of $\Gamma_{\mu}$ and $P_{\mu}$, the symmetric chiral rotation
is given by
\begin{equation}
\Gamma_{5}=\Gamma_{1}\Gamma_{2}\Gamma_{3}\Gamma_{4}P_{1}P_{2}P_{3}P_{4}
=(\gamma_{5}\otimes \gamma_{5})
=
\left(
\begin{array}{cc}
\gamma_{5}&   \\
 & -\gamma_{5}  \\
\end{array}
\right),
\end{equation}
\begin{align}
&\Psi_{n}\,\,\,\,\to\,\,\,\,\exp[i\,\theta\, (\gamma_{5}\otimes\gamma_{5})]\Psi_{n}\,,\,\,\,\,\,\,\,\,
\\
&\bar{\Psi}_{n}\,\,\,\,\to\,\,\,\,\bar{\Psi}_{n}\exp[i\,\theta\,(\gamma_{5}\otimes\gamma_{5})]\,.
\end{align}
which is the same flavored-chiral symmetry as the staggered chiral symmetry.

We list up relevant symmetries of this 4D four-flavor fermion action (\ref{4D4S}):
\vspace{2mm}

1. Flavored-non-singlet chiral symmetry with $\Gamma_{5}=\gamma_{5}\otimes\gamma_{5}$, 
\vspace{2mm}

2. $\gamma_{5}$-hermiticity with $\Gamma_{5}=\gamma_{5}\otimes \gamma_{5}$,
\vspace{2mm}

3. Discrete flavor-rotation symmetry $SW_{4}$,
\vspace{2mm}

4. Flavored-P and -T invariances associated with $\gamma_{5}$,
\vspace{2mm}

5. $U(1)$ baryon symmetry.
\vspace{2mm}

The flavored-parity stands for
\begin{align}
&\Psi_{n_{0}, n_{j}}\,\,\to\,\,i(\gamma_{5}\otimes \gamma_{4} )\Psi_{n_{0},-n_{j}}\,,
\\
&\bar{\Psi}_{n_{0}, n_{j}}\,\,\to\,\,-\bar{\Psi}_{n_{0},-n_{j}}i(\gamma_{5}\otimes \gamma_{4} )\,.
\end{align}
The flavored-time-reversal transformation is given by
\begin{align}
&\Psi_{n_{0},n_{j}}\,\,\to\,\,i(\gamma_{5}\otimes \gamma_{4}\gamma_{5} )\Psi_{-n_{0},n_{j}}\,,
\\
&\bar{\Psi}_{n_{0}, n_{j}}\,\,\to\,\,-\bar{\Psi}_{-n_{0},n_{j}}i(\gamma_{5}\otimes \gamma_{5}\gamma_{4} )\,.
\end{align}

This action could be useful for the $\mathcal{N}=4$ super Yang-Mills lattice formulation.
In the 4D $\mathcal{N}=4$ Sugimo model \cite{Su3}, 
the two-flavor structure with $8\times 8$ matrices for $\Gamma_{\mu}$ and $P_{\mu}$ is adopted.
In this setup, surplus modes other than $p=(0,0,0,0)$ emerge since
eight $8\times 8$ matrices $\Gamma_{\mu}$ and $P_{\mu}$
never fully anti-commute with one another.
Our present setup for the four-flavor lattice fermion adopts sixteen-spinor fermion fields 
with 8D $16\times 16$ gamma matrices,
which does not yield surplus modes with keeping exact chiral symmetry.
To construct the $\mathcal{N}=4$ SYM lattice model with four supercharges and four fermions,
our setup could be more proper.
The question is whether we can find the four-dimensional $\mathcal{N}=4$ lattice SYM action
which contains the following fermion action,
\begin{align} 
S_{F}^{\mathcal{N}=4}=-{1\over{4g_{0}^{2}}}\sum_{n,\mu}{\rm tr}
\Psi_{n}^{T}
[\Gamma_{\mu}(U_{n,\mu}\Psi_{n+\mu}-U_{n,-\mu}\Psi_{n-\mu})
+P_{\mu}(2-U_{n,\mu}\Psi_{n+\mu}-U_{n,-\mu}\Psi_{n-\mu})
]\,,
\end{align}
where $\Psi=(\psi_{A}, \psi_{B}, \psi_{C}, \psi_{D})^{T}$ is the four-flavor multiplet assembled into
a single 8D sixteen-spinor field.
and $\Gamma_{\mu}$ and $P_{\mu}$ belong to the 8D gamma matrices as (\ref{G4D4F}).
Further study is required to answer this question. 

In this section we discussed the four-flavor 4D extension 
of the pseudo higher-dimensional clifford-algebra method.
On the other hand, for practical application to lattice QCD, it is better if we 
reduce species (flavors) to two, the ``minimal" number.
In the next section, we propose the two-flavor 4D extension 
by introducing proper modification of the higher-dimensional clifford-algebra method.


\section{Extension to 4D two-flavor fermion}
\label{sec:4D}

In this section, we propose another way of extending the higher-dim clifford 
algebra method to four dimensions and obtain a two-flavor lattice fermion setup with exact chiral symmetry.
We first introduce a single-$P$ formulation.

\subsection{Single-$P$ formulation}
\label{sec:SP}

We consider the following 4D two-flavor lattice fermion action,
\begin{equation}
S={1\over{2}}\sum_{n}\sum_{\mu=1}^{4}\bar{\Psi}_{n}\left[  
\Gamma_{\mu}(U_{n,\mu}\Psi_{n+\mu}-U_{n,-\mu}\Psi_{n-\mu})
\,+\,iP(2-U_{n,\mu}\Psi_{n+\mu}-U_{n,-\mu}\Psi_{n-\mu})
\,+\, 2m_{0}\Psi_{n}
\right]\,,
\label{Sfinal}
\end{equation}
where a doublet fermion field $\Psi=(\psi_{A}, \psi_{B})^{T}$ 
is treated as a pseudo-6D eight-spinor.
$\Gamma_{\mu}$ and $P$ are $8\times 8$ hermite gamma matrices 
which are five out of six elements of 6D gamma matrices 
$\gamma_{M}^{(6)}$ ($M=1,\cdot\cdot\cdot, 6$).
We thus have the following relation
\begin{equation}
\{ \Gamma_{\mu}, \Gamma_{\nu}  \}=2\delta_{\mu\nu}\,, \,\,\,\,\,\,\,\,\,\,
\{ \Gamma_{\mu}, P \} = 0\,.
\label{ACfinal}
\end{equation}
In this modified formulation, we introduce a common gamma matrix $P$ 
for four-dimensional components of the Wilson term.
This modification, which we name the single-$P$ formulation, bypasses the problem discussed in (\ref{4cli-NA}). 

The associated Dirac operator in a free theory is given by
\begin{equation}
D(p)=i\Gamma_{\mu}\sin p_{\mu} +iP\sum_{\mu=1}^{4}(1-\cos p_{\mu}) + m_{0}\,.
\label{D4D}
\end{equation} 
Due to the anti-commuting relation (\ref{ACfinal}),
the denominator of the associated Dirac propagator,
\begin{equation}
|D(p)|^{2} =\sum_{\mu}^{4}\sin^{2} p_{\mu}+\left[\sum_{\mu}^{4}(1-\cos p_{\mu})\right]^{2} + m_{0}^2\,.
\label{P4D}
\end{equation}
has only one zero at $p=(0,0,0,0)$ for $m_{0}\to 0$.
This denominator of the propagator is exactly the same as that of the Wilson fermion.
Therefore, all the doublers, for example at $p=(\pi,0,0,0)$, have $O(1/a)$ mass, 
and are decoupled in the continuum limit.
The fermion action avoids further species-doubling and contains only two flavors, 
which are originally incorporated by the pseudo 8-spinor fermion field $\Psi$.

Consider the case we choose $\gamma^{(6)}_{1\sim5}$ for $\Gamma_{\mu}$ and $P$.
We then have two extra anti-commuting elements $\gamma^{(6)}_{6}$ 
and $\gamma^{(6)}_{7}=\gamma_{1}^{(6)}\gamma_{2}^{(6)}\cdot\cdot\cdot\gamma_{6}^{(6)}$.
It means that the action can possess larger subgroup of exact chiral symmetry, 
which is subgroup of $SU(2)$ axial symmetry.
We denote these two extra elements as $\Gamma_{5}^{\rm A}$ and $\Gamma_{5}^{\rm B}$.

In this case the recipe is given as follows,
\vspace{2mm}

(1) Assemble two flavors of 4D four-spinors into a single pseudo 6D eight-spinor field spuriously.
\vspace{2mm}

(2) Introduce five out of six 6D gamma matrices and assign four of them to
a kinetic term as $\Gamma_{\mu}$ and one to a Wilson term as $P$.
\vspace{2mm}

(3) Due to part of the 6D clifford algebra, we can avoid further species-doubling
with keeping subgroup of exact chiral symmetry.
\vspace{2mm}

The question here is whether we can find explicit forms of $\Gamma_{\mu}$ and $P$ 
to keep four-dimensional requisite symmetries, including flavor symmetry,
hypercubic symmetry, and C, P, T invariance.
(We note that there are lots of equivalent choices.
The point is just relative difference between $\Gamma_{\mu}$ and $P$.)

We consider the following explicit $\Gamma_{\mu}$ and $P$,
\begin{align}
\Gamma_{j}&={\bf 1}\otimes\sigma_{1}\otimes \sigma_{j}
={\bf 1}\otimes \gamma_{j}=\left(
\begin{array}{cc}
\gamma_{j} &  \\
 & \gamma_{j}  \\
\end{array}
\right)\,,\,\,\,\,\,\,\,\,\,\,(j=1,2,3),
\nonumber\\
\Gamma_{4}&={\bf 1}\otimes\sigma_{2}\otimes {\bf 1} 
={\bf 1}\otimes \gamma_{4}
=\left(
\begin{array}{cc}
\gamma_{4} &  \\
 & \gamma_{4}  \\
\end{array}
\right)\,,
\nonumber\\
P&=\sigma_{3}\otimes\sigma_{3}\otimes {\bf 1}
=\sigma_{3}\otimes \gamma_{5}
=\left(
\begin{array}{cc}
 \gamma_{5} &  \\
& -\gamma_{5}  \\
\end{array}
\right)\,.
\label{Gfinal}
\end{align} 
The free Dirac operator is given by
\begin{align}
D(p)&=i\Gamma_{\mu}\sin p_{\mu}+iP\sum_{\mu=1}^{4}(1-\cos p_{\mu}) +m_{0}
\nonumber\\
&=i({\bf 1}\otimes \gamma_{\mu})\sin p_{\mu}+i(\sigma_{3}\otimes \gamma_{5})\sum_{\mu}^{4}(1-\cos p_{\mu}) +m_{0}\,.
\label{Dfinal}
\end{align}
It is clear that the second terms of (\ref{Dfinal}) is the
Wilson term twisted by $i\sigma_{3}\otimes \gamma_{5}$.
We note that, even if we take $\sigma_{j}\otimes \gamma_{5}$ ($j=1,2$) as P, 
the structure does not change.

\subsection{Equivalence to Wilson Twisted-mass fermion}
\label{sec:Wtm}

The action (\ref{Dfinal}) is exactly the ``physical basis" representation of 
the Wilson full-twisted-mass fermion (Wtm fermion) \cite{F1, FR1, Ao1, GL1, Si1, Sh1, Adams3}.
The standard Wtm ``twisted-mass basis" is easily obtained
by changing the fermion variables \cite{Sh2} as
\begin{align}
&\Psi_{n}\,\,\to\,\,\Psi_{n}'\equiv\exp[i{\pi\over{4}}(\sigma_{3}\otimes\gamma_{5})]\Psi_{n}\,,\,\,\,\,
\\
&\bar{\Psi}_{n}\,\,\to\,\,\bar{\Psi}_{n}'\equiv\bar{\Psi}_{n}\exp[i{\pi\over{4}}(\sigma_{3}\otimes\gamma_{5})]\,.
\end{align}
The Dirac operator (\ref{Dfinal}) is then transformed to
\begin{equation} 
D(p)=i({\bf 1}\otimes \gamma_{\mu})\sin p_{\mu}-\sum_{\mu}(1-\cos p_{\mu}) +m_{0}(\sigma_{3}\otimes\gamma_{5})\,,
\label{DWtm}
\end{equation}
which is the well-known Wilson twisted-mass action.
As we mentioned, we have a variety of choices for $\Gamma_{\mu}$ and P as we will discuss in the end of this section.
Though, all the forms can be transformed into the Wtm fermion action by the change of variables.
In this sense, the present two-flavor lattice fermion action obtained here
is equivalent to the Wilson twisted-mass fermion action.

To show the details, we discuss symmetries of the action (\ref{Sfinal})(\ref{D4D}).
As shown above, we have two more hermite matrices anti-commuting $\Gamma_{\mu}$ and $P$.
We denote these two as $\Gamma_{5}^{\rm A}$ and $\Gamma_{5}^{\rm B}$,
which satisfy the relation,
\begin{equation}
\{ \Gamma_{\mu}, \,\Gamma_{5}^{\rm A, B}  \}=0\,, \,\,\,\,\,\,\,\,\,\,
\{ P, \,\Gamma_{5}^{\rm A, B} \} = 0\,,  \,\,\,\,\,\,\,\,\,\,
\{ \Gamma_{5}^{\rm A},\, \Gamma_{5}^{\rm B} \} = 0\,.
\end{equation}
This relation indicates the anti-commuting relation between $D(p)$ and $\Gamma_{5}^{\rm A,B}$
\begin{equation}
\{ D(p), \Gamma_{5}^{\rm A, B}  \}=0\,.
\end{equation}
It means that the lattice fermion action (\ref{Sfinal})(\ref{D4D}) in the massless limit $m_{0}\to 0$ 
is invariant under the subgroup of $SU(2)$ axial rotation
associated with $\Gamma_{5}^{\rm A}$ and $\Gamma_{5}^{\rm B}$
\begin{align}
&\Psi_{n}\,\,\,\,\to\,\,\,\,\exp[i\,\theta\, \Gamma_{5}^{\rm A,B}]\Psi_{n}\,,\,\,\,\,\,\,\,\,
\\
&\bar{\Psi}_{n}\,\,\,\,\to\,\,\,\,\bar{\Psi}_{n}\exp[i\,\theta\,\Gamma_{5}^{\rm A,B}]\,.
\end{align}
In the above explicit example (\ref{Dfinal}), $\Gamma_{5}^{\rm A,B}$ are given by
\begin{align}
\Gamma_{5}^{\rm A}&= 
\sigma_{1}\otimes\sigma_{3}\otimes {\bf 1}
=\sigma_{1}\otimes \gamma_{5}=\left(
\begin{array}{cc}
& \gamma_{5}   \\
 \gamma_{5} &  \\
\end{array}
\right)\,,
\nonumber\\
\Gamma_{5}^{\rm B}&= 
\sigma_{2}\otimes\sigma_{3}\otimes {\bf 1}
=\sigma_{2}\otimes \gamma_{5}=\left(
\begin{array}{cc}
 & -i\gamma_{5}  \\
i\gamma_{5} &  \\
\end{array}
\right)\,.
\end{align}
These are flavor-non-singlet chiral symmetries associated with $\sigma_{1}$ and $\sigma_{2}$,
which are subgroup of $SU(2)$ axial symmetry.
The other flavor-non-singlet chiral symmetry associated with $\sigma_{3}$ is broken
due to the Wilson-like term in (\ref{Sfinal})(\ref{Dfinal}) at $O(a)$.
This broken part causes additive renormalization of the operator 
$i\bar{\Psi}_{n}(\sigma_{3}\otimes\gamma_{5})\Psi_{n}$ and requires fine-tuning of the parameter.

We look into other lattice symmetries.
To investigate flavor symmetry, 
we again use the explicit basis (\ref{Dfinal}).
It has one exact flavor symmetry associated with $\sigma_{3}$ as,
\begin{align}
&\Psi_{n}\,\,\to\,\,\exp[i\,\theta\, (\sigma_{3}\otimes {\bf 1})]\Psi_{n}\,,\,\,\,\,\,\,\,\,
\\
&\bar{\Psi}_{n}\,\,\to\,\,\bar{\Psi}_{n}\exp[-i\,\theta\,(\sigma_{3}\otimes {\bf 1})]\,.
\end{align}
This is subgroup of $SU(2)$ vector symmetry.
We note the generator for this flavor symmetry is represented 
as $\sim i\Gamma_{5}^{\rm A}\Gamma_{5}^{\rm B}$.
We also have flavor-singlet vector symmetry, or $U(1)$ baryon symmetry.
\begin{align}
&\Psi_{n}\,\,\to\,\,\exp[i\theta ({\bf 1}\otimes {\bf 1})]\Psi_{n}\,,\,\,\,\,\,\,\,\,
\\
&\bar{\Psi}_{n}\,\,\to\,\,\bar{\Psi}_{n}\exp[-i\theta({\bf 1}\otimes {\bf 1})]\,.
\end{align}
Other flavor-non-singlet vector symmetries associated with $\sigma_{2}$ and $\sigma_{3}$ are broken by
the Wilson-like term in (\ref{Sfinal})(\ref{Dfinal}) at $O(a)$.
The total amount of flavor-non-singlet vector and axial symmetries does not change
in the twisted and non-twisted bases.
The total $SU(2)$ group is just transformed into each other as follows
\begin{equation}
[\sigma_{1}\otimes {\bf 1}]_{V}\,\otimes \,[\sigma_{2}\otimes {\bf 1}]_{V}\,\otimes\,[\sigma_{3}\otimes {\bf 1}]_{V}
\,\,\,\leftrightarrow\,\,\, 
[\sigma_{1}\otimes \gamma_{5}]_{A}\,\otimes \,[\sigma_{2}\otimes \gamma_{5}]_{A}\,\otimes\,[\sigma_{3}\otimes {\bf 1}]_{V}\,.
\end{equation}
where the left one is symmetry of the non-twisted basis while the right is that of the twisted basis.

The action (\ref{Dfinal}) has flavored-parity 
and flavored-time-reversal invariances.
The flavored-parity transformations are given by
\begin{align}
&\Psi_{n_{0}, n_{j}}\,\,\to\,\,i(\sigma_{1,2}\otimes \gamma_{4} )\Psi_{n_{0},-n_{j}}\,,
\\
&\bar{\Psi}_{n_{0}, n_{j}}\,\,\to\,\,-\bar{\Psi}_{n_{0},-n_{j}}i(\sigma_{1,2}\otimes \gamma_{4} )\,.
\end{align}
The flavored-time-reversal transformations are given by
\begin{align}
&\Psi_{n_{0},n_{j}}\,\,\to\,\,i(\sigma_{1,2}\otimes \gamma_{4}\gamma_{5} )\Psi_{-n_{0},n_{j}}\,,
\\
&\bar{\Psi}_{n_{0}, n_{j}}\,\,\to\,\,-\bar{\Psi}_{-n_{0},n_{j}}i(\sigma_{1,2}\otimes \gamma_{5}\gamma_{4} )\,.
\end{align}
Here we omit the transformation for the link variable for simplicity.
The usual parity and time-reversal are broken due to the Wilson-like term at $O(a)$.

The fermion action (\ref{Sfinal}) has the same hypercubic symmetry as Wilson fermion.
We then expect that Lorentz symmetry recovers in the continuum limit
even in the interacting theory.

We now list up all the relevant exact symmetries of our two-flavor fermion action:
\vspace{2mm}

1. Flavor-non-singlet chiral symmetry with $\Gamma_{5}^{\rm A}$ and $\Gamma_{5}^{\rm B}$, 
\vspace{2mm}

2. Flavor symmetry associated with $i\Gamma_{5}^{\rm A}\Gamma_{5}^{\rm B}$, 
\vspace{2mm}

3. Hypercubic symmetry,
\vspace{2mm}

4. Flavored-parity and flavored-time-reversal associated with $\sigma_{2}$ and $\sigma_{3}$,
\vspace{2mm}

5. $U(1)$ baryon symmetry.
\vspace{2mm}

These symmetries are the same as those of the Wilson twisted-mass action in a massless limit.
It exhibits that the present two-flavor lattice formulation is equivalent to the Wilson twisted-mass fermion.

For the different choices of $\Gamma_{\mu}$ and $P$, although the explicit forms of chiral, flavor and P, T
symmetries change, we essentially have the same symmetries which look different in a different basis.   
The associated action can be transformed to (\ref{Dfinal}) by appropriate change of variables.
For example, a different choice of $\Gamma_{\mu}$ and $P$ is given by
\begin{align}
\Gamma_{j}&={\bf 1}\otimes\sigma_{1}\otimes \sigma_{j}
={\bf 1}\otimes \gamma_{j}=\left(
\begin{array}{cc}
\gamma_{j} &  \\
 & \gamma_{j}  \\
\end{array}
\right)\,,\,\,\,\,\,\,\,\,\,\,(j=1,2,3),
\nonumber\\
\Gamma_{4}&=\sigma_{3}\otimes\sigma_{2}\otimes {\bf 1} 
=\sigma_{3}\otimes \gamma_{4}
=\left(
\begin{array}{cc}
\gamma_{4} &  \\
 & -\gamma_{4}  \\
\end{array}
\right)\,,
\nonumber\\
P&=\sigma_{1}\otimes\sigma_{2}\otimes {\bf 1}
=\sigma_{1}\otimes \gamma_{4}
=\left(
\begin{array}{cc}
& \gamma_{4}  \\
\gamma_{4} &  \\
\end{array}
\right)\,.
\label{G4DI}
\end{align}
In this basis, two chiral generators are transformed into
\begin{align}
\Gamma_{5}^{\rm A}&= 
{\bf 1}\otimes\sigma_{3}\otimes {\bf 1}
={\bf 1}\otimes \gamma_{5}=\left(
\begin{array}{cc}
\gamma_{5} &  \\
 & \gamma_{5}  \\
\end{array}
\right)\,,
\nonumber\\
\Gamma_{5}^{\rm B}&= 
\sigma_{2}\otimes\sigma_{2}\otimes {\bf 1}
=\sigma_{2}\otimes \gamma_{4}=\left(
\begin{array}{cc}
 & -i\gamma_{4}  \\
i\gamma_{4} &  \\
\end{array}
\right)\,.
\end{align}
It seems that this basis breaks hypercubic symmetry, while it keeps
flavored hypercubic symmetry.  
The flavor symmetry also seems to be broken at $O(1)$ in this basis.
These are all due to the change of basis.
The following change of the fermion variables transforms the associated action to (\ref{Dfinal}),
\begin{align}
\Psi\,\to\,\Psi' &\equiv \exp[i{\pi\over{4}}(\sigma_{3}\otimes \gamma_{4}\gamma_{5})]\exp[i{\pi\over{4}}({\bf 1}\otimes \gamma_{4}\gamma_{5})]
\exp[i{\pi\over{4}}(\sigma_{2}\otimes {\bf 1})]\Psi\,,
\\
\bar{\Psi}\,\to\,\bar{\Psi}'&\equiv \bar{\Psi}\exp[-i{\pi\over{4}}(\sigma_{2}\otimes {\bf 1})] \exp[-i{\pi\over{4}}({\bf 1}\otimes \gamma_{4}\gamma_{5})]\exp[-i{\pi\over{4}}(\sigma_{3}\otimes \gamma_{4}\gamma_{5})]\,.
\end{align}

Another seemingly-different basis is given by
\begin{align}
\Gamma_{j}&=\sigma_{3}\otimes\sigma_{1}\otimes \sigma_{j}
=\sigma_{3}\otimes \gamma_{j}=\left(
\begin{array}{cc}
\gamma_{j} &  \\
 & -\gamma_{j}  \\
\end{array}
\right)\,,\,\,\,\,\,\,\,\,\,\,(j=1,2,3),
\nonumber\\
\Gamma_{4}&=\sigma_{3}\otimes\sigma_{2}\otimes {\bf 1} 
=\sigma_{3}\otimes \gamma_{4}
=\left(
\begin{array}{cc}
\gamma_{4} &  \\
 & -\gamma_{4}  \\
\end{array}
\right)\,,
\nonumber\\
P&=\sigma_{1}\otimes{\bf 1}\otimes {\bf 1}
=\sigma_{1}\otimes {\bf 1}
=\left(
\begin{array}{cc}
& {\bf 1}  \\
{\bf 1} &  \\
\end{array}
\right)\,.
\label{G4DII}
\end{align}
In this choice, the two chiral generators are translated into
\begin{align}
\Gamma_{5}^{\rm A}&= 
\sigma_{3}\otimes\sigma_{3}\otimes {\bf 1}
=\sigma_{3}\otimes \gamma_{5}=\left(
\begin{array}{cc}
\gamma_{5} &  \\
 & -\gamma_{5}  \\
\end{array}
\right)\,,
\nonumber\\
\Gamma_{5}^{\rm B}&= 
\sigma_{2}\otimes{\bf 1}\otimes {\bf 1}
=\sigma_{2}\otimes {\bf 1}=\left(
\begin{array}{cc}
 & -i{\bf 1}  \\
i{\bf 1} &  \\
\end{array}
\right)\,.
\end{align}
In this case, it seems that hypercubic symmetry is 
intact while the flavor symmetry is broken at $O(1)$.
The following change of the fermion field transforms the action to (\ref{Dfinal}),
\begin{align}
\Psi\,\to\,\Psi' &\equiv \exp[i{\pi\over{4}}(\sigma_{3}\otimes \gamma_{5})]\exp[i{\pi\over{4}}({\bf 1}\otimes \gamma_{5})]
\exp[i{\pi\over{4}}(\sigma_{1}\otimes {\bf 1})]\Psi\,,
\\
\bar{\Psi}\,\to\,\bar{\Psi}'&\equiv \bar{\Psi}\exp[-i{\pi\over{4}}(\sigma_{1}\otimes {\bf 1})] \exp[-i{\pi\over{4}}({\bf 1}\otimes \gamma_{5})]\exp[-i{\pi\over{4}}(\sigma_{3}\otimes \gamma_{5})]\,.
\end{align}

We discuss feedback to the SUSY lattice formulation.
One of drawbacks in the Sugino model in four-dimensional $\mathcal{N}=2$ SYM \cite{Su3} is 
that it contains surplus fermionic modes as with the $\mathcal{N}=4$ SYM lattice model.
Our two-flavor lattice fermion has no surplus modes with keeping two exact chiral symmetries.
By applying our idea to the two-flavor Majorana fermion actions, we may have
a better lattice formulation of $\mathcal{N}=2$ SYM. 
Unlike the case of application of the halved-translation-invariance fermion to $\mathcal{N}=4$ SYM in the previous section,
the single-P formulation or Wtm do not mix rotation and flavor subscripts 
and do not suit the twisted-SUSY algebra in the Sugino model. 
Thus, we first need to find $\mathcal{N}=2$ lattice SUSY algebra, which incorporates the Wilson twisted-mass fermion
in the associated action.


\section{Summary}
\label{sec:SUM}

In this paper, we investigate the fermion action of the two-dimensional $\mathcal{N}=(2,2)$ super Yang-Mills lattice 
model and discuss its extension to general cases including four-dimensional lattice fermions.
We mainly found out the following two points.

(1) The reconcilement of exact chiral symmetry and absence of further species-doubling
in the 2D $\mathcal{N}=(2,2)$ SYM lattice formulation originates in
the spurious 4D four-spinor identification of 2D two-flavor fields with the four-dimensional 
clifford algebra embedded into the action.

(2) This fermion construction by use of the higher-dimensional clifford algebra is 
extended to four-dimensional lattice fermions in two manners:
In the halved-translation-invariance fermion, a 4D four-flavor multiplet is spuriously identified
as a pseudo 8D sixteen-spinor with 8D gamma matrices introduced, 
then further species-doubling is bypassed without breaking of whole chiral symmetry.
The other setup, which was shown to be equivalent to the Wilson twisted-mass fermion, 
treats a 4D two-flavor multiplet as a pseudo 6D eight-spinor
with part of 6D gamma matrices introduced into the action, and realizes the lattice fermion with two species, 
subgroup of chiral symmetry, hypercubic symmetry and flavored-P, T invariances.

In Sec.~\ref{sec:SUSY} we review the lattice Sugino model of two-dimensional 
${\mathcal N}=(2,2)$ SYM and investigate the fermion action part.
In the formulation, the two Majorana flavors are assembled into a single pseudo four-spinor field,
with four 4D gamma matrices assigned to the kinetic and Wilson-like terms.
Due to this structure, the associated Dirac propagator has no doubler pole, 
thus the number of species never exceeds the minimum number, two (two Majorana spinors in this case).
The main point is that the lattice fermion action incorporates the higher-dimensional clifford algebra structure, 
and chiral-$U(1)_{R}$ symmetry is realized as the pseudo higher-dimensional chiral symmetry.

In Sec.~\ref{sec:2D}, as a natural extension of the Sugino fermion action, 
we discuss the 2D Dirac lattice fermion with two flavors and exact flavored-chiral symmetry.
By rewriting the action into the flavor-spin representation, 
we find its similarities to and differences from staggered fermions.
The main differences are the flavor symmetry breaking at $O(a)$ and 
the halved translation invariance per dimension \cite{La1}.
We term the action Halved-translation-invariance fermion.

In Sec.~\ref{sec:4D4} we first show that the naive application of the higher-dimensional clifford algebra method 
to 4D lattice fermions leads to surplus modes as shown in \cite{Su3}.
We propose a way of extending the structure to 4D Dirac fermions by starting with a four-flavor multiplet field:
We spuriously treats 4D four flavors as a single 16-spinor field with
introducing eight 8D gamma matrices into the action.
The embedded 8D clifford algebra avoids further species-doubling with keeping subgroup of exact chiral symmetry.
We show that the action is a four-dimensional version of the halved-translation-invariance fermion.
We discuss if the fermion setup can
contribute to lattice formulation of four-dimensional $\mathcal{N}=4$ SYM theory.

In Sec.~\ref{sec:4D}, we discuss 4D two-flavor lattice fermion actions:
In the action, a 4D Dirac doublet field is treated as a pseudo 6D 8-spinor field 
with five of six 6D gamma matrices incorporated into kinetic and Wilson terms.
The fermion action has no further species-doubling and possesses two exact chiral symmetries. 
This two-species lattice fermion also keeps hypercubic symmetry
and flavored-P,T invariances, which is shown to be equivalent to the Wilson twisted-mass fermion.

Lastly we discuss application of the fermion actions to SYM on the lattice.
The halved-translation-invariance fermion with four flavors and chiral-$U(1)_{R}$ symmetry 
could contribute to improvement of the 4D $\mathcal{N}=4$ SYM lattice formulation.
The two-flavor formulation with a single $P$, or the Wilson twisted-mass fermion, 
can be a criterion for constructing 4D $\mathcal{N}=2$ SYM lattice theory. 
Further study on the higher-dim clifford algebra formulation of lattice fermions
may lead to a new class of practically and theoretically valuable lattice setups.


\begin{acknowledgments}
T.M. is grateful to JICFuS (Joint Institute for Computational Fundamental Science) 
and the organizers for making an opportunity for him to give a talk on and discuss the present subject.
He is thankful to T.~Onogi for helping find a final form of the two-flavor chiral fermion.
He appreciate kind instruction on the Wilson twisted-mass fermion given by A.~Shindler and M.~Creutz.
He thanks S.~Aoki, T~Kimura, S.~Matsuura, H.~Sugino and D.~Kadoh for the fruitful discussion.
\end{acknowledgments}


\end{document}